\documentstyle[aps,pre]{revtex}
\newcommand{\eff}{{{\mathrm eff}}}
\begin{document} 
 
\title{Ponderomotive force of {\em quasi-particles} in a plasma} 
 
\author{L.O. Silva$^{\P}$,R.Bingham$^{\S}$, J.M.Dawson$^{\P}$, 
 W.B.Mori$^{\P}$} 
\address{$^{\P}$Department of Physics and Astronomy \\ 
University of California  Los Angeles, Los Angeles, California 90095} 
\address{$^{\S}$Rutherford Appleton Laboratory \\
Chilton, Didcot, Oxon OX11 OQX, U.K.}
 
\draft 
\maketitle 
 
\begin{abstract} 
We derive the force exerted on the background plasma  
 by an arbitrary distribution of non interacting {\em 
quasi-particles},  
corresponding to either collective excitations of the plasma  
(plasmons, phonons) or {\em dressed particles}  
 (photons, neutrinos). Our approach is based on the effective   
 Hamiltonian describing the quasi-classical dynamics of the 
individual  
 particles in the presence of a background medium. We recover the 
usual  
 results for the relativistic ponderomotive force of a photon gas, 
and we derive the  
force, due to weak interactions, exerted by the electron-neutrinos in 
a  
background medium containing electrons, positrons and neutrons with 
arbitrary distribution functions. Generalization to other background 
species 
and other neutrino flavors is also discussed.   
\end{abstract} 
  
\section{Introduction} 
 
The ponderomotive force of electromagnetic waves 
\cite{kibble,schmidt,lindman,manheimer,mckinstrie,mora}
is a key concept in plasma physics \cite{kruer,chen} and plays a 
central role in our   
present understanding of intense laser-plasma interactions 
\cite{liukaw}.  
This force arises whenever a non-uniform oscillating  
electric field is present in a  
dielectric, and can be seen as a slow time scale effect, or the 
average effect, 
due to some non-uniformity of the high frequency oscillations  
of the electric field \cite{landau}.  
 
In general, the derivation of the ponderomotive force is based on the 
analysis  
of the single particle dynamics of charged particles in the presence 
of the  
electric field 
\cite{kibble,schmidt,lindman,manheimer,mckinstrie,mora} 
or in terms of the Maxwell equations for a macroscopic media 
\cite{landau}. 
By averaging the motion of the charged particle over the  
fast time scale, corresponding to the high frequency oscillations  
of the field,  
the slow time scale dynamics of the individual particles can then be 
calculated,  
and the net effect of the electromagnetic forces  
acting on the particle can be reduced to the ponderomotive force.  
In this paper, we present  
 a different approach, which allows us not only to rederive the 
previous  
results,  
but also to generalize, in a straightforward way, the concept of 
ponderomotive force 
 to other physical conditions or other {\em quasi-particles} besides 
photons 
(e.g. {\em dressed} neutrinos), where the dynamics of a single 
electron is 
 not easily described in terms of the classical force in the 
relativistic equations 
 of motion. 
 
The starting point of our treatment is a semi-classical description 
of the  
 fields interacting with the plasma. The fields are described by  
their equivalent {\em quasi-particles}, or elementary quanta:  
photons for the electromagnetic field, plasmons  
for the longitudinal electrostatic oscillations, phonons for the ion 
acoustic oscillations,  
and dressed neutrinos for the neutrino field interacting with the  
background medium.  
This is done by defining a distribution function for the particles 
from the 
field intensity. Such a description is very useful because a kinetic 
equation 
 can be derived for the correct quasi-particles distribution. The 
kinetics or 
equations of motion for the 
  {\em quasi-particles} in the plasma are described  
by the dispersion relation, or the effective Hamiltonian, for each 
one of  
the fields or {\em quasi-particles}. The effective Hamiltonian can be 
derived either from a classical or a quantum field theory. 
The key point of our formalism is that by knowing the effective 
Hamiltonian  
 describing a single {\em quasi-particle} dynamics,   
we are then able to derive the force exerted by a gas  
of non interacting elementary quanta in the background plasma:  
 the dispersion relation, 
or the effective Hamiltonian, generates the equations of motion of 
the quasi-particles  
 and from conservation of momentum (action-reaction) it generates the 
force of 
 the quasi-particles in the plasma. 

Even if the term  
{\em ponderomotive force} was coined to describe the forces  
acting on a dielectric in  
an arbitrary non uniform electric field, in this paper 
 we generalize the concept to the interaction of  
any non uniform field with a background medium. In fact, the force 
exerted in a background medium  
 due to a non uniform field can also be seen as the pressure gradient 
arising due  
to some inhomogeneity in the quasi-particle distribution. 
% A 
% similar picture was proposed by Zeldovich and Novikov 
% \cite{zeldovich}, 
% where two types of gravitational pressure were identified: 
% the usual field pressure associated with the mean energy density of 
% the field, 
% and a second type of "pressure" associated with nonuniformities or 
% fluctuations 
% of the gravitational field.
 
 We aim to achieve two objectives: to derive the ponderomotive force 
solely 
based on the quasi-particle concept, thus providing an easy tool to 
generalize and 
unify the concept for different physical scenarios, and to explicitly 
show 
the relation between the ponderomotive force and the wave action 
density, 
or the quasi-particle number density. 
%------ Added in the revised version ------------
In particular, we will be mainly 
 interested in determining the ponderomotive force of an arbitrary 
distribution of neutrinos 
(with different flavors) in an arbitrary background medium.
%------------------------------------------------

Our focus will be on the derivation of the force exerted by an 
arbitrary distribution of  
 {\em quasi-particles}. In Section II, we present our formalism and 
the approximations  
 involved in our description. The general expression for the 
ponderomotive force is then derived  
 and represented as a function of the effective Hamiltonian or 
dispersion relation
 (describing the quasi-classical
 dynamics of the {\em quasi-particles}) and the {\em quasi-particles} 
distribution function. 
In Section III, we apply our results to classical fields  
arising in a plasma. 
The ponderomotive force exerted by  
 a gas of photons in a plasma is rederived in the relativistic 
regime, 
 thus showing the equivalence between our new approach  
 and  the derivation of the ponderomotive force found in the 
literature.
We also derive the  force due to a gas of plasmons and a  
gas of phonons. In Section IV, we consider the  
interaction of a gas of neutrinos with a dense plasma, also 
containing other species  
of baryonic matter (such as neutrons, protons and positrons). The 
ponderomotive force 
exerted in the  
background medium by the neutrinos 
 is then derived. In our derivation we assume that each one of the 
species in the  
medium has an arbitrary distribution function, thus generalizing 
previous results only valid for  
 cold plasmas \cite{bingham2}. 
Finally, in Section V, the results of this paper are summarized.

\section{General Considerations} 
 
Consider a gas of non interacting {\em quasi-particles} (QPs) in a 
background medium.  
By {\em quasi-particles}, we mean not only elementary collective 
excitations of the background medium,  
such as plasmons and phonons, but also dressed photons or dressed 
neutrinos. Our formalism is  
independent of the entities we are considering as QPs, but we assume 
that the  
interaction {\bf between} them is negligible. This corresponds to 
either  
considering collision frequencies $\nu_{qq}$ much smaller than the 
typical time scale of  
the process in study(dilute gas) or simply assuming that the 
interaction between quasi-particles can be  
discarded. Furthermore, we consider that the dynamics of a single QP 
is  
governed by the effective Hamiltonian $H_{\eff}$, which is a function 
not only of the dynamical  
variables of the single QP (momentum and position)  
 but is also dependent of the properties of the medium where the QPs 
propagate. 
 
 Using in a direct way the results already known for the 
ponderomotive force  
of e.m.waves in a dielectric in terms of a gradient in the laser 
intensity, 
 we could immediately propose an expression for the 
ponderomotive force, written as a function of the number of photons. 
This approach would not show, however, the connection between the 
effective interaction felt by a single QP and the expression of the 
ponderomotive force, thus preventing the generalization to other 
physical conditions.
Also, the dependence of the ponderomotive force on the wave action 
density, or QP 
number density, would not be clearly stated for non trivial QP 
distribution functions.
Therefore, a different path must be followed. By determining the free 
energy of the 
system, we can relate the changes in the free energy 
with the work performed by the force exerted by the distribution of 
particles, and 
from that derive the ponderomotive force.

We can write the total free energy as the sum of two contributions: 
the  
free energy of the background medium in the absence of QPs ($F_0$)  
and the additional 
free energy (for a given temperature and density) resulting from the 
presence  
of the field elementary excitations. Therefore, the free energy is 
written as  
\begin{equation} 
F=F_0(\rho,T)+g_{sq} \int d{\mathbf r} \int \frac{d{\mathbf k}}{(2 
\pi)^3} f_q({\mathbf r},{\mathbf k},t) H_{\eff}  
\label{eq:free} 
\end{equation} 
where $\rho$ is the density of the medium, $T$ is the temperature of 
the medium, and  
$f_q$ describes the QP distribution function  in phase-space  
(${\mathbf r},{\mathbf k}$), ${\mathbf r}$ describes the QP position 
and ${\mathbf k}$ is related to the 
QP momentum ${\mathbf p}_q$ by ${\mathbf k}={\mathrm  p}_q/\hbar$. 
 For the sake of completeness we state here the most important 
properties of $f_q$: 
\begin{equation} 
 n_q({\mathbf r},t)= g_{sq} \int \frac{d{\mathbf k}}{(2 \pi)^3} 
f_q({\mathbf r},{\mathbf k},t) \label{eq:p1} \quad , 
\end{equation} 
\begin{equation} 
 |\psi({\mathbf r},t)|^2= g_{sq} \int \frac{d{\mathbf k}}{(2 \pi)^3} 
f_q({\mathbf r},{\mathbf k},t)  H_{\eff} \label{eq:p2} \quad , 
\end{equation} 
\begin{equation} 
 |\phi({\mathbf k},t)|^2= g_{sq} \int d{\mathbf r} f_q({\mathbf 
r},{\mathbf k},t)  H_{\eff} \label{eq:p3} \quad , 
\end{equation} 
where $n_q$ is the {\em quasi-particle} number density,  
$|\psi({\mathbf r},t)|^2$ is the  
 spatial energy density, and $|\phi({\mathbf k},t)|^2$ is the 
spectral energy density, so that  
the second term on the right hand side of equation (\ref{eq:free}) 
represents the total free energy  
due to the presence of QPs. $g_{sq}$ is the QP statistical weight and 
it accounts for 
spin degeneracy of the QPs.
 
We now generalize the procedure described in ref.\cite{landau} 
 to electromagnetic waves.
Let us assume that an isothermal deformation ${\mathbf u}$  
of the background infinite medium occurs.  
If the distribution of {\em quasi-particles} exerts a force ${\mathbf 
f}$ over the medium,   
work $\delta W$ will be done by that force, such that 
\begin{equation} 
\delta W = \int  d{\mathbf r} \, {\mathbf f} \cdot {\mathbf u} \quad 
. 
\end{equation}  
 where ${\mathbf f}$ has the dimensions of force per unit volume.  
We consider that this force will act only on the electrons of the 
background medium. The ponderomotive 
force felt by the ions can be discarded since it is $m_e/m_i$ smaller 
than the ponderomotive force felt 
by the electrons ($m_e$ is the electron rest mass, and $m_i$ is the 
ion mass). 
 However, inclusion of other background species is straightforward. 
$\delta W$ can be related with the change  
 in the free energy for the same displacement ${\mathbf u}$, since 
$\delta W=-\delta (F-F_0)$.  
 Going back to eq.(\ref{eq:free}), we easily obtain 
\begin{equation} 
\delta F = \delta F_0(\rho,T)+g_{sq} \int d{\mathbf r} \int 
\frac{d{\mathbf k}}{(2 \pi)^3} 
 \left( H_{\eff} \delta f_q({\mathbf r},{\mathbf k},t) +f_q({\mathbf 
r},{\mathbf k},t) \delta H_{\eff} \right)  
 \label{eq:dfree} 
\end{equation} 
We will assume a quasi-static distribution of  
 {\em quasi-particles}, which means that a displacement of the 
background medium will not  
 affect the distribution $f_q$. Furthermore, $f_q$ is a function 
only  
 of the dynamical variables. Hence, the second term 
%added in revised manuscript
 in the integral  
 in eq.(\ref{eq:dfree}) is zero.  
 The effective energy $H_{\eff}$ of each QP will be affected by  
the isothermal background deformation, since $H_{\eff}$ depends not 
only on the dynamical variables  
but also on the properties of the background medium. The change in 
$H_{\eff}$ has two  
contributions: (i) $\delta_{(1)} H_{\eff}$, due to the fact that 
particles from the background  
medium are pushed from ${\mathbf r}-{\mathbf u}$ to ${\mathbf r}$, 
and (ii) $\delta_{(2)} H_{\eff}$,   
due to the change of the distribution function, or density, of the 
background medium  
in position ${\mathbf r}$. The first contribution for $\delta H_\eff$ 
is  
\begin{equation} 
 \delta_{(1)} H_{\eff}=-{\mathbf u} \cdot \nabla H_{\eff} 
\label{eq:dw1} 
\end{equation} 
For the remaining contribution, we first use the fact that the 
relative change in the  
volume element of the background medium is $dV/V=\nabla \cdot 
{\mathbf u}$.  
Therefore, the change  
in the number density $n_{\mathrm bg}$ is $\delta n_{\mathrm bg} = - 
n_{\mathrm bg}\nabla \cdot {\mathbf u}$, 
 where $n_{\mathrm bg}$ is the number  
density of the background particles affected by the ponderomotive 
force (in our case, just the electrons).
In the same way,  
the change in the distribution function of the background medium 
particles  
$f_{{\mathrm bg}}({\mathbf r},{\mathbf p})$ is  
$\delta f_{{\mathrm bg}}({\mathbf r},{\mathbf p},t)=-f_{{\mathrm 
bg}}({\mathbf r},{\mathbf p})  
\nabla \cdot {\mathbf u}$, where ${\mathbf p}$ is the momentum of the 
background particles
 (we also assume that the isothermal deformation does not impart 
momentum to the medium).  
Therefore, $\delta_{(2)} H_\eff$ can be written either as  
\begin{equation} 
 \delta_{(2)} H_{\eff}= -\left(\frac{\partial H_\eff}{\partial  
n_{\mathrm bg}} \right)_T  n_{\mathrm bg} \nabla  
 \cdot {\mathbf u} \label{eq:dw2a} 
\end{equation} 
if $H_\eff$ is a function of $ n_{\mathrm bg}$, or 
\begin{equation} 
 \delta_{(2)} H_{\eff}= - \left(\frac{\partial H_\eff}{\partial  
 f_{{\mathrm bg}}} \right)_T f_{{\mathrm bg}}({\mathbf r},{\mathbf 
p}) \nabla  
 \cdot {\mathbf u}  \label{eq:dw2b} 
\end{equation} 
 for a more general dependence of $H_\eff$ on the distribution 
function of the  
background medium $f_{{\mathrm bg}}$. Inserting eq.(\ref{eq:dw1}) and 
eq.(\ref{eq:dw2a}),  
 in eq.(\ref{eq:dfree}) we obtain 
\begin{equation}  
\delta F -\delta F_0(\rho,T)=- g_{sq} \int d{\mathbf r} \int 
\frac{d{\mathbf k}}{(2 \pi)^3} f_q({\mathbf r},{\mathbf k},t) 
 \left({\mathbf u} \cdot \nabla H_\eff +\left(\frac{\partial 
H_\eff}{\partial  n_{\mathrm bg}} \right)_T  
 n_{\mathrm bg} \nabla  
 \cdot {\mathbf u} \right) \label{eq:dfree2} 
\end{equation} 
Performing an integration by parts over the second term on the 
right-hand side of  
 eq.(\ref{eq:dfree2}), results in 
\begin{eqnarray}  
 \delta F -\delta F_0(\rho,T)=- g_{sq} \int d{\mathbf r} \int 
\frac{d{\mathbf k}}{(2 \pi)^3} {\mathbf u} \cdot  
 f_q({\mathbf r},{\mathbf k},t) \nabla H_\eff + \nonumber  
 \\  g_{sq} \int d{\mathbf r} \int \frac{d{\mathbf k}}{(2 \pi)^3} 
{\mathbf u} \cdot \nabla  
 \left(f_q({\mathbf r},{\mathbf k},t) \left( \frac{\partial H_\eff}  
 {\partial  n_{\mathrm bg}} \right)_T  n_{\mathrm bg} \right) 
\end{eqnarray} 
Therefore, the ponderomotive force, per unit volume, acting on the 
medium is 
\begin{equation} 
{\mathbf f}({\mathbf r},t)=g_{sq} \int \frac{d{\mathbf k}}{(2 \pi)^3} 
 f_q({\mathbf r},{\mathbf k},t) \nabla H_\eff -g_{sq} \int 
\frac{d{\mathbf k}}{(2 \pi)^3} \nabla  
 \left(f_q({\mathbf r},{\mathbf k},t) \left( \frac{\partial H_\eff}  
 {\partial  n_{\mathrm bg}} \right)_T  n_{\mathrm bg} \right) 
\label{eq:pondf} 
\end{equation} 
A similar expression can also be derived when $H_\eff$ depends on the 
distribution function of  
the background medium $f_{{\mathrm bg}}$. This will be discussed for 
the particular case of the  
ponderomotive force due to neutrinos in a plasma.  
 For a linear dependence of $H_\eff$ on $n_{\mathrm  bg}$ {\it i.e.}  
$n_{\mathrm bg} (\partial H_\eff/\partial n_{\mathrm bg})=H_\eff$, 
eq.(\ref{eq:pondf}) can be further simplified to 
\begin{equation} 
{\mathbf f}({\mathbf r},t)=-g_{sq} \int \frac{d{\mathbf k}}{(2 
\pi)^3} H_\eff \nabla f_q({\mathbf r},{\mathbf k},t) 
\end{equation}  
 
Equation (\ref{eq:pondf}), and the appropriate definition of 
$H_\eff$,  
is the starting point of our discussion of the ponderomotive force  
due to different types of QPs propagating in a plasma. It also 
represents a 
generalization for arbitrary fields of the Landau and Lifshitz 
arguments \cite{landau}. 
  
\section{Ponderomotive force of photons, plasmons and phonons} 
 
Having determined the general expression for the ponderomotive force 
of a gas  
of non-interacting QPs, we now proceed by evaluating 
eq.(\ref{eq:pondf}) for  
different classical fields or classical QPs. 
We first consider the ponderomotive force due to a  
distribution of photons characterized by $f_q \equiv {\cal 
N}({\mathbf r},{\mathbf k},t)$. The number of  
photons $\cal{N}$, obeying the properties (\ref{eq:p1}-\ref{eq:p3})  
can be obtained from the Wigner function of the electromagnetic field 
\cite{silva1}.  
 The concept of the number of photons was introduced in the Plasma 
Physics literature in 
the 60's, associated with the Random Phase Approximation 
\cite{sagdeev}. 
We point out, however, that a 
proper definition of the number of photons based on the Wigner 
function, can describe 
any e.m.field configuration (see Appendix, for a detailed discussion 
of this problem).
 For the number of photons $\cal N$, the role of the effective 
Hamiltonian $H_\eff$  
 is played by $\hbar \omega({\mathbf r},{\mathbf k},t) \equiv \hbar  
\omega_{\mathrm k}$, 
 where the frequency $ \omega_{\mathrm k}$ is  
 obtained from the dispersion relation for the electromagnetic  
waves propagating in the plasma. This is equivalent to assuming that 
each single photon obeys the    
dispersion relation for plane electromagnetic waves. Since we are 
considering an arbitrary distribution of photons,  
the relativistic mass correction must also be included in the 
dispersion relation for  
circularly polarized electromagnetic plane waves: 
\begin{equation} 
 \omega_{\mathrm k}^2=k^2 c^2 + \frac{\omega_{pe}^2 ({\mathbf 
r},t)}{\gamma} \label{eq:ph_dis} 
\end{equation} 
where $\omega_{pe}^2({\mathbf r},t)=4 \pi e^2 n_e({\mathbf r},t)/m_e$ 
is the local electron plasma frequency,  
 and $n_e({\mathbf r},t)$ is the electron number density. $\gamma$ is 
the  
relativistic mass correction factor, which is a function of the 
electric field intensity and the  
 energy distribution function of the electrons in the background 
plasma. For the sake of clarity,  
we will not write down the explicit expression for $\gamma$, since it 
will not add any additional  
features to our derivation. We note that in (\ref{eq:ph_dis}), 
${\mathbf k}$, ${\mathbf r}$, and $t$ are  
independent variables. Furthermore, eq.(\ref{eq:ph_dis}) is valid in 
the limits of the  
geometrical optics approximation \cite{bernstein} {\it i.e.}   
\begin{equation} 
 \frac{\omega_{\mathrm k}}{2 \pi} \gg \frac{\partial}{\partial t} 
\log n_e({\mathbf r},t)\quad , \qquad 
 \frac{|{\mathbf k}|}{2 \pi} \gg \left|\nabla \log n_e({\mathbf 
r},t)\right|  
 \label{eq:wkb} 
\end{equation} 
These limits establish the conditions in which this approach is 
valid: 
whenever two very different time scales 
are present, we can treat the high frequency perturbations as a gas 
of 
quasi-particles propagating in a background with slow density 
modulations. 
 
Using eq.(\ref{eq:ph_dis}) in eq.(\ref{eq:pondf}) and the fact that 
\begin{equation} 
\left(\frac{\partial H_\eff}{\partial  n_{\mathrm bg}}\right)_T  
n_{\mathrm bg} \equiv  
 \left(\frac{\partial \hbar \omega_{\mathrm k}}{\partial 
\omega_{pe}^2}\right)_T \omega_{pe}^2 =  
 \frac{\hbar \omega_{pe}^2}{2 \omega_{\mathrm k} \gamma} 
\end{equation} 
and  
\begin{equation} 
 \nabla H_\eff \equiv \nabla \hbar \omega_{\mathrm k} = 
\frac{\hbar}{2 \omega_{\mathrm k}} \nabla 
\frac{\omega_{pe}^2}{\gamma} 
\end{equation} 
we obtain, after some algebra, the ponderomotive force due to a 
photon distribution $\cal N$: 
\begin{equation} 
{\mathbf f}_{\mathrm ph}({\mathbf r},t)=- \frac{\omega_{pe}^2}{2 
\gamma} g_{ph} \nabla \int  
 \frac{d {\mathbf k}}{(2 \pi)^3} \hbar \frac{{\cal 
N}}{\omega_{\mathrm k}} \label{eq:ph_pond} 
\end{equation} 
where $g_{ph}=1(2)$ for circular(linear) polarization.
For plane electromagnetic waves,  
 ${\cal N}=\frac{|E_0|^2}{8 g_{ph} \pi \hbar \omega_0} 
\delta({\mathbf k}-
 {\mathbf k}_0)$, with $E_0$ the  
electric field amplitude, and $\omega_0 \, (k_0)$ the frequency(wave 
number) of  
the electromagnetic field \cite{silva1}. In this case, the 
ponderomotive  
force acting on a single electron reduces to the more familiar form: 
\begin{equation} 
{\mathbf f}_{\mathrm ph}({\mathbf r},t)=-\frac{e^2}{2 m_e \gamma} 
\nabla {\mathbf A}^2 \label{eq:ph_pond2}
\end{equation} 
where ${\mathbf A}$ is the high frequency vector potential. It is 
straightforward to see  
 that ponderomotive force effects can arise due to two different 
conditions:  
 inhomogeneity in the number of photons distribution function $\cal 
N$ and/or  
 spatial dependent frequency. Present understanding of ultraintense 
short laser pulse propagation in 
plasmas is based on the different roles played by these two 
contributions \cite{mori}. 
Equation (\ref{eq:ph_pond2}) agrees with previous derivations of the 
relativistic ponderomotive force for 
circularly polarized photons 
\cite{kibble,schmidt,manheimer,mckinstrie,mora}. 
For linearly polarized plane electromagnetic waves, the dispersion 
relation 
(\ref{eq:ph_dis}) is no longer valid; a different expression for the 
relativistic ponderomotive 
force appears \cite{lindman}, which under certain conditions reduces 
to (\ref{eq:ph_pond2}).
As far as we know, eq.(\ref{eq:ph_pond}) is the first derivation of 
the relativistic ponderomotive force, 
using solely the number of photons concept.  Also, 
eq.(\ref{eq:ph_pond}) is valid 
for any e.m.field configuration, as long as ${\cal N}$ is defined 
using the proper 
definition of Wigner function for an arbitrary e.m.field (see 
Appendix).

For a gas of plasmons (electron plasma oscillations), an expression 
similar to eq.(\ref{eq:ph_pond}) can also be derived in a similar 
manner, but now the dispersion 
relation is 
\begin{equation} 
\omega_{\mathrm k}^2=\omega_{pe}^2+3 k^2 \frac{k_B T_e}{m_e} 
\end{equation} 
where $T_e$ is the plasma electron temperature.
The ponderomotive force of the gas of plasmons, described by 
${\cal N}_{\mathrm plasmon}$, acting 
 on the plasma electrons is 
\begin{equation} 
{\mathbf f}_{\mathrm pl}({\mathbf r},t)=- \frac{\omega_{pe}^2}{2} 
\nabla \int  
 \frac{d {\mathbf k}}{(2 \pi)^3} \hbar \frac{{\cal N}_{\mathrm 
plasmon}}{\omega_{\mathrm k}} \label{eq:pl_pond}
\end{equation} 
The ponderomotive force of the gas of plasmons can then describe the 
coupling of the 
electrostatic oscillations with the low frequency, long wavelength 
ion acoustic oscillations 
\cite{sagdeev}. 
 
We now consider the ponderomotive force due to a gas of 
elementary collective ion excitations. Due to 
the limits of our formalism (see eq.(\ref{eq:wkb})), this gas of ion 
oscillations 
 must interact with even lower 
frequency and longer wavelength plasma perturbations. In a 
unmagnetized 
electron-ion plasma, such physical condition cannot be verified. 
However, dusty plasmas 
support oscillations with characteristic frequencies(wavelengths) 
much lower(longer) than  
 those of ion oscillations (ion acoustic waves or ion plasma waves) 
\cite{shukla}. 
The picture of a gas of ion oscillations in a dust-acoustic 
oscillation is then reasonable. 
The general linear dispersion relation for ion oscillations is 
\cite{chen}
\begin{equation} 
 \omega_{\mathrm k}^2 = k^2 \frac{\gamma_i k_B T_i}{m_i}+ 
 k^2 \frac{\gamma_e k_B T_e}{m_i}\frac{1}{1+\gamma_e k^2 \lambda_e^2} 
\label{eq:dispion} 
\end{equation} 
with $T_i$ is the ion temperature, 
 $\gamma_e(\gamma_i)$ is the electron(ion) adiabatic index, 
 and $\lambda_e=(k_B T_e/ 4\pi n_e e^2)^{1/2}$ is the electron Debye 
length. 
Denoting the number density of QP excitations corresponding 
to ion collective motions by ${\cal N}_{\mathrm ion \, qp}$, 
we obtain from eq.(\ref{eq:pondf}), 
and using eq.(\ref{eq:dispion}), the force: 
\begin{equation} 
{\mathbf f}_{\mathrm ion \, qp} ({\mathbf r},t)=-\frac{m_i}{k_B T_e} 
\frac{\lambda_e^2}{2} 
\nabla \int  \frac{d {\mathbf k}}{(2 \pi)^3} 
 \frac{\hbar{\cal N}_{\mathrm ion \, qp} }{\omega_{\mathrm k}} 
\left(\frac{k^2 \gamma_e k_B T_e}{m_i} 
 \frac{1}{1+\lambda_e^2 k^2 \gamma_e} \right)^2 \label{eq:fion}
\end{equation} 
In the limit of $\gamma_i T_i/m_i \ll \gamma_e T_e/m_i$, the 
dispersion relation 
 (\ref{eq:dispion}) 
can be approximated by $\omega_{\mathrm k}^2 \simeq k^2 \gamma_e k_B 
T_e/\{m_i(1+ \lambda_e^2 k^2  \gamma_e)\}$, and 
eq.(\ref{eq:fion}) reduces to 
\begin{equation} 
 {\mathbf f}_{\mathrm ion \, qp}({\mathbf r},t)=-\frac{m_i}{k_B T_e} 
\frac{\lambda_e^2}{2} \nabla 
 \int  \frac{d {\mathbf k}}{(2 \pi)^3} \hbar{\cal N}_{\mathrm ion \, 
qp} 
 \omega_{\mathrm k}^3 
\end{equation} 

Two opposite physical scenarios can now be explored. In the limit of 
$k \lambda_e \ll 1$, 
corresponding to ion acoustic oscillations, the dispersion relation 
is 
$\omega_{\mathrm k} \simeq k c_s$, with the 
sound speed $c_s=\sqrt{\gamma_e k_B T_e/m_i}$. Thus, the 
ponderomotive force due to a distribution 
of ion acoustic waves, or a gas of ion acoustic phonons (iap), is 
\begin{equation}
 {\mathbf f}_{\mathrm iap}({\mathbf r},t)=-\frac{1}{2} 
\frac{c_s^3}{\omega_{pi}^2} \nabla 
  \int \frac{d {\mathbf k}}{(2 \pi)^3} \hbar k^3 {\cal N}_{\mathrm 
iap} \label{eq:f_iap}
\end{equation} 
with the ion plasma frequency $\omega_{pi}=\omega_{pe} 
\sqrt{m_e/m_i}$.
On the other hand, in the limit of $k \lambda_e \gg 1$, the 
dispersion relation 
(\ref{eq:dispion}) describes ion plasmas waves (ipw), such that 
 $\omega_{\mathrm k}^2 \simeq \omega_{pi}^2$. 
 In this case, the ponderomotive force verifies 
\begin{equation}
{\mathbf f}_{\mathrm ipw}({\mathbf r},t)=\omega_{pi} \nabla 
 \int  \frac{d {\mathbf k}}{(2 \pi)^3} \hbar {\cal N}_{\mathrm ipw}- 
 \frac{3}{2} \nabla  
 \int  \frac{d {\mathbf k}}{(2 \pi)^3} \hbar \omega_{\mathrm k} {\cal 
N}_{\mathrm ipw} 
\label{eq:f_ipw}
\end{equation}

We then see that the ponderomotive force associated with ion 
collective excitations, 
eqns.(\ref{eq:f_iap},\ref{eq:f_ipw}), is 
significantly different from that associated with photons, 
eq.(\ref{eq:ph_pond}), or plasmons, 
eq.(\ref{eq:pl_pond}). This derivation must now be verified using the 
standard plasma physics methods. 
The coupling between the ion collective motions and the background 
medium can play a 
significant role in the (de)stabilization of dust-acoustic 
oscillations in a dusty plasma. 
Both these features will be explored in a future work.
 
\section{Ponderomotive force of neutrinos in dense plasmas} 

We now turn to the central result of this paper: the ponderomotive 
force of neutrinos in a plasma. 
This force can provide the coupling mechanism responsible for the 
anomalous 
scattering of neutrinos and the consequent deposition in the plasma
of some part of the neutrino energy released in a supernovae 
explosion \cite{bingham2,bingham1}. 
This energy deposition plays a key role in the present understanding 
of supernovae 
explosions \cite{wilson}.
An intuitive picture of the ponderomotive force is immediate: if in a 
given region 
the energy density in the neutrinos is higher than in other regions, 
a force will be exerted in the background medium by the neutrinos, 
towards the regions of lower neutrino energy density. This 
corresponds to the physical picture of the neutrinos trying to push 
their way into 
regions of lower energy density.

In order to derive the ponderomotive force due to the 
electron-neutrinos ($\nu_e$) 
 propagating in a background medium we first write down the effective 
Hamiltonian for a single 
 electron-neutrino in a unmagnetized background of electrons 
characterized 
 by the electron distribution function $f_e({\mathbf r},{\mathbf 
p}_e,t)$ \cite{nunokawa}:
\begin{equation}
H_{\mathrm eff}=\sqrt{p_\nu^2 c^2+m_\nu^2 c^4}+g_{se} \int \frac{d 
{\mathbf p}_e}{(2 \pi)^3} 
 V_{\mathrm eff}({\mathbf r},{\mathbf p}_e,t) \label{eq:dispnu}
\end{equation}
with $V_{\mathrm eff}$ given by
\begin{equation}
V_{\mathrm eff}({\mathbf r},{\mathbf p}_e,t)=g_V \sqrt{2} G_F 
  f_e({\mathbf r},{\mathbf p}_e,t) 
 \left(1-\frac{{\mathbf p}_e \cdot \hat{\mathbf k}_\nu}{E_e}\right)  
\label{eq:veff}
\end{equation} 
where $G_F$ is the Fermi constant of weak interaction, 
 $g_V=(1/2+2 \sin^2 \theta_W) \simeq 1$ is the effective vector 
coupling constant in the Standard Model, 
 $\theta_W$ is the Weinberg mixing angle,  
 ${\mathbf p}_\nu=\hbar {\mathbf k}$ is the electron-neutrino 
momentum, 
 $\hat{{\mathbf k}}_\nu={\mathbf p}_\nu/|{\mathbf p}_\nu|$, $E_e$ is 
the electron energy 
(a function of the electron momentum ${\mathbf p}_e$), 
and $m_\nu$ is the neutrino mass, which can be set to zero for 
massless neutrinos. 
 $g_{se}=2$ is the statistical weight for the electrons, 
corresponding to spin $1/2$, and appears 
because of spin degeneracy.  
Also, $f_e({\mathbf r},{\mathbf p}_e,t)$ satisfies 
\begin{equation}
 n_e({\mathbf r},t)=g_{se} \int \frac{d {\mathbf p}_e}{(2 \pi)^3} 
f_e({\mathbf r},{\mathbf p}_e,t) 
 \label{eq:ne_def}
\end{equation} 

The effective potential in Eq.(\ref{eq:veff}) 
has been derived using the methods of Finite Temperature Quantum 
Field Theory \cite{ftft}. 
Hence, our approach here is 
clearly a semi-classical one: we assume that the interaction of the 
neutrinos with the electrons 
is governed by quantum processes (included in $V_{\mathrm eff}$), and 
we take into account the Fermi 
statistics of the phase space density of the particle numbers, but 
the neutrino dynamics is 
 determined by the classical Hamiltonian obtained using the 
equivalence 
principle, and we neglect the spins. 
This approximation is valid as long as changes in 
 $V_{\mathrm eff}$ occur over length scales much longer than 
 the neutrino de Broglie wavelength $\lambda_\nu= 2 \pi/|{\mathbf 
k}|$, and no spin waves are considered.  

We first analyze the contribution of the term ${\mathbf p}_e \cdot 
\hat{\mathbf k}_\nu/E_e$ to 
the effective Hamiltonian (\ref{eq:dispnu}). If the neutrinos 
propagate along a precise direction, 
and assuming an isotropic distribution for the electrons, the 
integral 
\begin{equation}
 \int \frac{d {\mathbf p}_e}{(2 \pi)^3} f_e({\mathbf r},{\mathbf 
p}_e,t)
\frac{{\mathbf p}_e \cdot \hat{\mathbf k}_\nu}{E_e} 
\end{equation}
 averages to zero. Furthermore, if the neutrino distribution is 
isotropic, when integrating 
over the contribution of all the neutrinos for the ponderomotive 
force, a zero average is 
obtained once again. Therefore, this term only gives a contribution 
for both anisotropic 
neutrino {\bf and} electron distribution functions. 
% In spite of its negligible contribution we will 
% keep this term in our derivation.
  
Since eq.(\ref{eq:dispnu}) is also a function of the distribution 
function of the 
electrons in the background medium, eq.(\ref{eq:dw2b}) must now be 
employed to obtain the 
expression for the ponderomotive force due to the neutrinos:
\begin{eqnarray}
 {\mathbf f}_\nu ({\mathbf r},t) & = & g_{se} \int \frac{d {\mathbf 
k}}{(2 \pi)^3} 
 \int \frac{d {\mathbf p}_e}{(2 \pi)^3} f_\nu({\mathbf r},{\mathbf 
k},t) 
 \nabla V_{\mathrm eff}- \nonumber \\
 & & - g_{se}\int \frac{d {\mathbf k}}{(2 \pi)^3} 
 \int \frac{d {\mathbf p}_e}{(2 \pi)^3} \nabla \left(f_\nu({\mathbf 
r},{\mathbf k},t) 
 \left(\frac{\partial V_{\mathrm eff}}{\partial f_e} \right)_T 
 f_e({\mathbf r},{\mathbf p}_e,t) \right) \label{eq:nu_pond1}
\end{eqnarray}
where $f_\nu({\mathbf r},{\mathbf k},t)$ is the neutrino distribution 
function, 
which also verifies the normalization condition
\begin{equation}
n_\nu({\mathbf r},t)=g_{s \nu} \int \frac{d {\mathbf k}}{(2 \pi)^3} 
 f_\nu({\mathbf r},{\mathbf k},t) \label{eq:nnu_def}
\end{equation}
where $n_\nu({\mathbf r},t)$ is the neutrino number density, and 
$g_{s \nu}=1$ is the 
neutrino statistical weight (neutrinos are completely polarized 
(left) particles). Since 
the effective potential has a linear dependence on $f_e({\mathbf 
r},{\mathbf p}_e,t)$, then 
\begin{equation}
\left(\frac{\partial V_{\mathrm eff}}{\partial f_e} \right)_T 
 f_e({\mathbf r},{\mathbf p}_e,t) = V_{\mathrm eff}
\end{equation}
 and eq.(\ref{eq:nu_pond1}) reduces to 
\begin{equation}
 {\mathbf f}_\nu ({\mathbf r},t)= - g_{se} \int \frac{d {\mathbf 
k}}{(2 \pi)^3} 
 \int \frac{d {\mathbf p}_e}{(2 \pi)^3} V_{\mathrm eff} \nabla 
 f_\nu({\mathbf r},{\mathbf k},t) \label{eq:nu_pond2}
\end{equation}
Inserting eq.(\ref{eq:veff}) in eq.(\ref{eq:nu_pond2}), we obtain
\begin{eqnarray}
{\mathbf f}_\nu ({\mathbf r},t)= - 
\frac{\sqrt{2}}{2} (1+4 \sin^2 \theta_W) G_F g_{se} \int \frac{d 
{\mathbf k}}{(2 \pi)^3} 
 \int \frac{d {\mathbf p}_e}{(2 \pi)^3} f_e({\mathbf r},{\mathbf 
p}_e,t) 
 \nabla f_\nu({\mathbf r},{\mathbf k},t) + \nonumber \\ 
 \frac{\sqrt{2}}{2}  (1+4 \sin^2 \theta_W) G_F g_{se} \int \frac{d 
{\mathbf k}}{(2 \pi)^3} 
 \int \frac{d {\mathbf p}_e}{(2 \pi)^3} f_e({\mathbf r},{\mathbf 
p}_e,t) 
 \frac{{\mathbf p}_e \cdot \hat{\mathbf k}_\nu}{E_e} 
 \nabla f_\nu({\mathbf r},{\mathbf k},t) \label{eq:nu_pond3}
\end{eqnarray}
Using eqns.(\ref{eq:ne_def},\ref{eq:nnu_def}), the first term in 
eq.(\ref{eq:nu_pond3}) 
 can be re-written as 
\begin{equation}
 {\mathbf f}_\nu ({\mathbf r},t)=- \frac{\sqrt{2}}{2} 
  (1+4 \sin^2 \theta_W) G_F n_e({\mathbf r},t) \nabla n_\nu({\mathbf 
r},t) 
 \label{eq:nu_pond}
\end{equation}
Eq.(\ref{eq:nu_pond3}) represents the force per unit volume exerted 
by the neutrinos 
over the electrons contained in the unit volume. The second term, 
which accounts for the 
 contributions from anisotropies, is different from zero only when 
the electron distribution function {\bf and} the neutrino 
distribution function are 
anisotropic. 
This means that even for a beamed neutrino distribution, the second 
term vanishes for an isotropic plasma. 
From now on we will consider an isotropic electron distribution 
function, 
 and hence we discard the contribution of the 
second term in eq.(\ref{eq:nu_pond3}).
The first term in eq.(\ref{eq:nu_pond3}) (or eq.(\ref{eq:nu_pond})) 
is then responsible 
 for the strong coupling between 
the neutrinos and the electrons, as suggested before \cite{bingham1}. 
From eq.(\ref{eq:nu_pond}), we can easily 
derive the force acting on a single electron due to the presence of 
the neutrino 
distribution:
\begin{equation}
 {\mathbf f}_{\nu - {\mathrm e}}=- \frac{\sqrt{2}}{2} 
  (1+4 \sin^2 \theta_W) G_F \nabla n_\nu({\mathbf r},t) 
\label{eq:nu_pondsingle}
\end{equation} 
We stress out that Eq.(\ref{eq:nu_pond}) is valid for any neutrino or 
electron distribution functions.

The generalization of 
Eq.(\ref{eq:nu_pond3},\ref{eq:nu_pond},\ref{eq:nu_pondsingle}) 
 for the force exerted by neutrinos and 
antineutrinos over electrons, positrons or the neutrons is also 
straightforward. Changing the effective potential 
$V_{\mathrm eff}$\cite{kuo} in order to include the weak interaction 
of the electron-neutrinos and anti electron-neutrinos 
with electrons(positrons), we obtain the ponderomotive force over a 
single 
electron(positron) 
\begin{equation}
 {\mathbf f}_{\nu \, \bar{\nu}-e^+(e^-)} = \mp \frac{\sqrt{2}}{2} 
(1+4 \sin^2 \theta_W)  G_F \nabla 
 \left(n_{\nu e}({\mathbf r},t)-n_{\bar{\nu}e}({\mathbf r},t) \right)
\end{equation}
where the minus(plus) sign refers to electrons(positrons),  
being $n_{\bar{\nu}e}({\mathbf r},t)$ the anti electron-neutrino 
number density. 
The ponderomotive force over a single neutron can also be written as
\begin{equation}
 {\mathbf f}_{\nu \, \bar{\nu}-n}=\frac{\sqrt{2}}{2} G_F \nabla 
 \left(n_{\nu e}({\mathbf r},t)-n_{\bar{\nu}e}({\mathbf r},t) \right)
\end{equation}
As before, we are assuming a unmagnetized background medium, and we 
are discarding the 
contribution of the anisotropies of the electron/positron/neutron and 
neutrino/antineutrino 
distribution functions. Generalization of the ponderomotive force due 
to other neutrino 
 flavors (tau-neutrinos or muon-neutrinos) is straightforward as long 
as the proper effective 
potentials $V_\eff$ are considered \cite{kuo}.

It is now important to compare our expression of the ponderomotive 
force with that first 
introduced by Bingham {\it et al} \cite{bingham2} in a 
phenomenological way, 
or the one derived by Hardy and Melrose \cite{hardy} using the 
methods of 
Quantum Plasmadynamics \cite{qpd}. 
The expression derived by Bingham {\it et al} is based on the analogy 
between the 
ponderomotive force due to electromagnetic waves (as derived in 
\cite{landau}) 
and the ponderomotive force due to the neutrinos, and it is only 
valid 
 in the limits of validity of the Landau and Lifshitz expression {\it 
i.e.} 
as long as the neutrino flux is assumed monochromatic and the energy 
(or frequency) of 
the neutrinos is assumed constant. In this particular physical 
scenario eq.(11) 
in ref.\cite{bingham2} is equivalent to our eq.(\ref{eq:nu_pond}).
 A comparison with the results derived in ref.\cite{hardy} shows that 
their results are 
equivalent to those presented here, in the limit of constant 
background electron/positron 
number density. The discrepancy is present whenever gradients of the 
background number 
density are assumed, and it arises from a misinterpretation of the 
ponderomotive force concept  
 in reference \cite{hardy}. A 
correct interpretation of the results obtained by the Quantum 
Plasmadynamics formalism 
gives the same expression for the ponderomotive force of 
electron-neutrinos 
in a background of electrons and positrons as the one derived here 
\cite{silva2}.

It must be stressed that a long range interaction force between 
neutrinos and electrons 
was identified before in the context of a quantum kinetic treatment 
of neutrinos in 
 a lepton plasma. 
An equivalent expression to eq.(\ref{eq:nu_pond}) is evident in 
eq.(18') 
of ref.\cite{semikoz}. The long range interaction 
force is associated with an effective charge the neutrinos acquire in 
a 
background of electrons \cite{semikoz,nucharge}.
Our results allow us to clearly identify the long range force in 
\cite{semikoz} 
as the ponderomotive force due to the weak interaction of the 
neutrinos with a plasma. 

\section{Summary} 
 
In this paper, we have presented a general derivation of the 
ponderomotive force due to quasi-particles 
propagating in a background medium. In particular, the results for 
the relativistic ponderomotive force of 
photons and for the ponderomotive force due to plasmons were 
recovered. The force of a gas of 
ion collective motions in the background plasma was also presented. 
We then applied the same techniques 
to a gas of neutrinos interacting with a background medium through 
the weak interaction force. The ponderomotive 
force due to an arbitrary distribution of neutrinos and antineutrinos 
interacting either with electrons, positrons or neutrons was derived.

The ponderomotive force derived here is the force on a Lagrangian 
fluid element or on a single particle from 
the background medium. It then 
provides the proper way to build a single-particle or a 
self-consistent 
kinetic theory for the interaction of QPs (neutrinos, photons, 
plasmons, phonons) 
with a background medium, 
which can be applied as the foundation tool for the study of QPs 
driven instabilities 
in a plasma. 

\bigskip
This work was partially supported 
by NSF Contracts AST-9713234 and DMS-9722121 and DOE grants 
DE-FG03-98DP00211 and DE-FG03-92ER40727. LOS acknowledges the 
financial 
 support of PRAXIS XXI through grant BPD11804/97.

\section{Appendix}
\renewcommand{\theequation}{A.\arabic{equation}}
\setcounter{equation}{0}

In order to clarify the meaning and scope of the number of photons 
concept, we 
present a calculation of the number of photons ${\cal N}$ for 
several e.m.field configurations. This will allow us to connect 
this concept with previous definitions found in the literature and to 
 clearly point out that no Random Phase Approximation is assumed
when a proper definition for the number of photons is employed. 

The number of photons ${\cal N}$ obeys the general properties 
described by 
eqns.(\ref{eq:p2},\ref{eq:p3}) for the distribution function $f_q$. 
The class of phase-space distribution functions 
that verify these properties for a given wave field, are usually 
denoted as 
Wigner functions \cite{wigner}. 
The most common representation of the Wigner function for  
the electric field  ${\mathbf E}({\mathbf r},t)$ is
\begin{equation}
{\cal N}({\mathbf k},{\mathbf r},t)= \frac{1}{8 g_{ph} \pi \hbar 
\omega_{\mathbf k}
 ({\mathbf r},t)}  
 \int d {\mathbf s} 
  {\mathbf E}({\mathbf r}-{\mathbf s}/2,t) \cdot {\mathbf 
E}^*({\mathbf r}+{\mathbf s}/2,t)
 \exp(i {\mathbf k} \cdot {\mathbf s}) 
\end{equation}
where $\omega_{\mathbf k}({\mathbf r},t)$ is obtained from the 
dispersion relation 
$D(\omega,{\mathbf k},{\mathbf r},t) \equiv 0$. For the wave field 
described 
by ${\cal N}$, equation (\ref{eq:p1}) 
defines the wave action density, and eq.(\ref{eq:p2}) defines the 
field energy density.
The Wigner function exactly satisfies a kinetic equation which 
reduces to a Vlasov 
equation in the short wavelength high frequency approximation.

For simplicity, we will consider 
propagation in vacuum, thus meaning that $\omega_{\mathbf 
k}=|{\mathbf k}|c$. We 
now calculate ${\cal N}$ for different electric fields. We first 
start with a 
plane wave monochromatic wave, ${\mathbf E}({\mathbf r},t) = {\mathbf 
E}_0 \exp i ({\mathbf k}_0 \cdot 
 {\mathbf r}-\omega_0 t)$, being $\omega_0$ the frequency of the 
electric field and 
${\mathbf k}_0$ the wavevector. The number of photons is then simply 
written as 
\begin{equation}
{\cal N}_{\mathrm plane}({\mathbf k},{\mathbf r},t)= 
\frac{|{\mathbf E}_0|^2}{8 g_{ph} \pi \hbar 
 |{\mathbf k}| c} 
 \delta({\mathbf k}-{\mathbf k}_0) 
\end{equation}
describing a monochromatic beam of photons, as expected. This is also 
the usual 
result present in the Plasma Physics literature since the 60's 
\cite{tsytovich}. 
 A test of the ability of the Eq.(A.1) for the definition of the 
 number of photons ${\cal N}$ to describe the 
 complete structure of the 
electric field arises whenever the electric field depicts an 
interference 
 pattern. The most simple case corresponds to an electric field 
described by 
the superposition of two plane waves
${\mathbf E}({\mathbf r},t) = {\mathbf E}_0 \exp i ({\mathbf k}_0 
\cdot 
 {\mathbf r}-\omega_0 t)+{\mathbf E}_1 \exp i ({\mathbf k}_1 \cdot 
 {\mathbf r}-\omega_1 t)$. The number of photons for this electric 
field 
 defined by Eq.(A.1) is 
\begin{eqnarray}
 {\cal N}_{\mathrm BW}({\mathbf k},{\mathbf r},t) & = & \frac{1}{8 
g_{ph} \pi \hbar 
 |{\mathbf k}| c} \left[ |{\mathbf E}_0|^2 \delta({\mathbf 
k}-{\mathbf k}_0)+
 |{\mathbf E}_1|^2 \delta({\mathbf k}-{\mathbf k}_1)+ 
\right.\nonumber \\ 
 & & \left. {\mathbf E}_0 \cdot 
 {\mathbf E}_1^* \cos ({\bf \Delta}_k \cdot {\mathbf r}-\Delta_\omega 
t) 
 \delta\left({\mathbf k}-\frac{{\mathbf k}_0+{\mathbf 
k}_1}{2}\right)\right] 
\end{eqnarray}
where ${\bf \Delta}_k={\mathbf k}_1-{\mathbf k}_0$, and 
$\Delta_\omega=
 \omega_1-\omega_0$. 
The presence of two photon beams, associated with the two plane waves 
is obvious. 
Furthermore, a third beam is present which results from the 
interference between 
the two plane waves, showing the characteristic slow modulation of 
the beat pattern. 
Using eq.(\ref{eq:p2}), we 
can immediately recover the usual energy density for two 
 interfering monochromatic planes waves.
It is then clear, that the number of 
photons, as described from eq.(A.1), already contains the information 
about any 
interference pattern present in the electric field. 

To clarify this point, and 
make a bridge to the usual definitions, we consider the 
superposition of several monochromatic plane waves, also containing 
an additional 
random phase factor $\psi_{\mathbf k}$. The electric field is then 
written as 
${\mathbf E}({\mathbf r},t)= \sum_{{\mathbf k}'} 
{\mathbf A}_{k'} \exp i({\mathbf k}' \cdot {\mathbf r} - 
\omega_{{\mathbf k}'} t+\psi_{{\mathbf k}'})$. For this electric 
field, eq.(A.1) 
 verifies 
\begin{eqnarray}
{\cal N}_{\mathrm RP}({\mathbf k},{\mathbf r},t) & = &
 \frac{1}{8 g_{ph} \pi \hbar |{\mathbf k}| c} \sum_{{\mathbf k}'} 
\sum_{{\mathbf k}''} 
 {\mathbf A}_{{\mathbf k}'} {\mathbf A}^*_{{\mathbf k}''} \nonumber \\
 & & \times \exp i( {\bf \Delta}_{{\mathbf k}' {\mathbf k}''} \cdot 
{\mathbf r}
  - {\Delta}_{\omega ' \omega ''} t +
 {\Delta}_{\psi' \psi''}) \delta({\mathbf k}-\frac{{\mathbf k}'
+{\mathbf k}''}{2}) 
\end{eqnarray}
 where ${\bf \Delta}_{{\mathbf k}' {\mathbf k}''} = {\mathbf k}'- 
{\mathbf k}'' $, 
 ${\Delta}_{\omega ' \omega ''} = \omega '- \omega ''$, 
 and  ${\Delta}_{\psi' \psi''}=\psi_{{\mathbf k}'}- \psi_{{\mathbf 
k}''}$.
Once again, the beat interference pattern observed in eq.(A.3) is 
also present for 
the terms verifying ${\bf \Delta}_{{\mathbf k}' {\mathbf k}''} \not=  
0$ and 
 ${\Delta}_{\omega ' \omega ''} \not= 0$.
Eq.(A.4) is the generalization of eq.(A.3) for the superposition of 
an arbitrary 
 number plane waves. 
We stress that so far, no assumptions have been made regarding the 
properties 
of the phases $\psi_{\mathbf k}$. When the phases $\psi_{\mathbf k}$ 
are random, a phase averaging of eq.(A.4) can be performed. This 
averaging 
corresponds to the well known Random Phase Approximation (RPA) 
\cite{tsytovich}. 
Being the phase averaging $\equiv <> $ 
defined as the average of a statistical ensemble of systems 
 differing from one another only in the phase $\psi_{{\mathbf k}}$ or 
 ${\Delta}_{\psi' \psi''}$, it is obvious that 
$< \exp i {\Delta}_{\psi' \psi''}>= \delta({\mathbf k}' -{\mathbf 
k}'')$, 
thus leading to the RPA number of photons distribution function
\begin{eqnarray}
{\cal N}_{\mathrm RPA}({\mathbf k},{\mathbf r},t) & = &
 <{\cal N}_{\mathrm RP}({\mathbf k},{\mathbf r},t)> = 
 \frac{1}{8 g_{ph} \pi \hbar |{\mathbf k}| c} \sum_{{\mathbf k}'} 
\sum_{{\mathbf k}''} 
 {\mathbf A}_{{\mathbf k}'} {\mathbf A}^*_{{\mathbf k}''} \nonumber \\
 & & \times \exp i( {\bf \Delta}_{{\mathbf k}' {\mathbf k}''} \cdot 
{\mathbf r}
 - {\Delta}_{\omega ' \omega ''} t ) 
 \delta ({\mathbf k}' -{\mathbf k}'') 
 \delta({\mathbf k}-\frac{{\mathbf k}'+{\mathbf k}''}{2})  \nonumber 
\\
 & = & \frac{1}{8 g_{ph} \pi \hbar |{\mathbf k}| c} \sum_{{\mathbf 
k}'} 
 |A_{{\mathbf k}'}|^2 \delta({\mathbf k}-{\mathbf k}') 
\end{eqnarray}
This is the conventional definition of the number of photons, only 
valid 
under the limits of the Random Phase Approximation. No interference 
pattern is present, thus describing  independent and non interfering 
 photon beams.

From this discussion, it becomes evident that a definition of the 
number of 
photons based on the Wigner function can rigorously describe  
different e.m.field 
configurations, including those where interference between different 
field components 
is important. In the limit of the Random Phase Approximation, the 
usual  
definitions are recovered. It may be argued that the Wigner function 
presents 
some pathologies (it is not a positive-definite function) 
which, at first sight, could prevent its use. However, we stress 
that the quantities with straightforward 
 physical meaning are the marginals of the Wigner 
function, eqns.(\ref{eq:p1},\ref{eq:p2},\ref{eq:p3}), and these 
possess the 
correct physical properties.


\begin{thebibliography}{99}
\bibitem{kibble} T.W.Kibble, Phys.Rev.Lett. {\bf 16}, 1054 
\bibitem{schmidt} G.Schmidt and T.Wilcox, Phys.Rev.Lett. {\bf 31}, 
1380 (1973)
\bibitem{lindman} E.L.Lindman and M.A.Stroscio, Nucl.Fusion {\bf 17}, 
619 (1977)
\bibitem{manheimer}  W.M.Manheimer, Phys.Fluids {\bf 28}, 1569 (1985)
\bibitem{mckinstrie} C.J.McKinstrie and D.F.Dubois, Phys.Fluids {\bf 31}, 
 278 (1988); E.A.Startsev and C.J.McKinstrie, Phys.Rev. E {\bf 55}, 7527 (1997).
\bibitem{mora} P.Mora and T.M.Antonsen Jr., Phys.Plasmas {\bf 4} 217 
(1997).
\bibitem{kruer} W.L.Kruer, {\it The Physics of Laser Plasma 
Interactions} 
 (Addison-Wesley, Redwood City, CA, 1988).
\bibitem{chen} F.F.Chen, {\it Introduction to Plasma Physics}, 2nd 
Ed. (Plenum, New York, 1984).
\bibitem{liukaw} C.S.Liu and P.K.Kaw in {\it Advances in Plasma 
Physics} Vol. 6 
Eds. P.K.Kaw, W.L.Kruer, C.S.Liu and K.Nichikawa, (Wiley, New York, 
1986) p.83.
\bibitem{landau} L.Landau, E.M.Lifshitz and L.P.Pitaevsky,  
 {\it Electrodynamics of Continuous Media} 2nd Ed. 
(Butterworth-Heinemann, Oxford, 1995).
% \bibitem{zeldovich} Ya.B.Zeldovich and I.D.Novikov, 
% {\it Stars and Relativity}, 
% (Dover, New York, 1996), pp.155-157.
\bibitem{bingham2} R.Bingham, H.A.Bethe, J.M.Dawson, {\it et al.}, 
 {\it Phys.Lett. A} {\bf 220}, 107 (1996).
% \bibitem{tajima} T.Tajima and K.Shibata, {\it Plasma Astrophysics}, 
%  (Addison-Wesley, Reading MA, 1997), pp.450-451. 
% \bibitem{nodar} N.L.Tsintsadze, J.T.Mendon\c{c}a, L.N.Tsintsadze, 
%  submitted to Phys.Plasmas (1998).
\bibitem{silva1} L.O. Silva and J.T.Mendon\c{c}a, Phys.Rev. E {\bf 
57}, 
 3423 (1998).
\bibitem{sagdeev} R.Z.Sagdeev and A.A.Galeev, {\it Nonlinear Plasma 
Theory} 
 (W.A.Benjamin, New York, 1969).
\bibitem{bernstein} I.B.Bernstein and L.Friedland, in {\it Handbook 
of Plasma 
Physics, Vol 1:Basic Plasma Physics I}, Eds. A.A.Galeev and R.N.Sudan 
(North 
Holland, New York, 1983), p.365.
\bibitem{mori} W.B.Mori, IEEE J.Quantum Electr. {\bf 33}, 1942 (1997).
\bibitem{shukla} N.N.Rao, P.K.Shukla, and M.Y.Yu, Planet.Space Sci. 
{\bf 38}, 
 543 (1990); P.K.Shukla and V.P.Silin, Phys.Scr. {\bf 45}, 504 (1992).
\bibitem{bingham1} R.Bingham, J.M.Dawson, J.J.Su, H.A.Bethe, 
Phys.Lett. A {\bf 193}, 
 279 (1994).
\bibitem{wilson} J.R.Wilson, in {\it Numerical Astrophysics}, 
ed.J.M.Centrella, 
J.M.leBlanc, R.L.Bowers (Jones \& Bartlett, Boston, 1985), p.422; 
H.A.Bethe and 
J.R.Wilson, Ap.J. {\bf 295}, 14 (1985).
\bibitem{nunokawa} H.Nunokawa, V.B.Semikoz, A.Y.Smirnov, and 
J.W.F.Valle, 
Nucl.Phys. B {\bf 501}, 17 (1997) and references therein.
\bibitem{ftft} L.Dolan and R.Jackiw, Phys.Rev. D {\bf 9}, 3320 (1974).
\bibitem{kuo} T.K.Kuo and J.Pantaleone, Rev.Mod.Phys. {\bf 61}, 937 
(1989).
\bibitem{hardy} S.J.Hardy and D.B.Melrose, Phys.Rev. D {\bf 54}, 6491 
(1996).
\bibitem{qpd} D.Melrose, Plasma Phys. {\bf 16}, 845 (1974).
\bibitem{silva2} L.O.Silva, R.Bingham, J.M.Dawson, {\it et al.}, 
physics/9807050. 
\bibitem{semikoz} V.B.Semikoz, Physica A {\bf 142}, 157 (1987); 
 V.N.Oraevsky and V.B.Semikoz, Sov.Phys. JETP {\bf 59}, 465 (1984).
\bibitem{nucharge} J.T.Mendon\c{c}a, L.O.Silva, R.Bingham {\it et 
al.}, Phys.Lett. A, 
 {\bf 239}, 373 (1997).
\bibitem{wigner} E.Wigner, Phys.Rev. {\bf 40}, 749 (1932).
\bibitem{tsytovich}V.N.Tsytovich, {\it Nonlinear Effects in Plasma}, 
(Plenum Press, New York, 1970).
\end{thebibliography}
\end{document}